\documentclass[reprint,twocolumn,prb,amsmath,amssymb,aps]{revtex4-2} 

\usepackage{graphicx}  
\usepackage{dcolumn}   
\usepackage{bm}        
\usepackage{hyperref}  
\hypersetup{colorlinks=true, linkcolor=blue!95!black!85!yellow, citecolor=blue!95!black!85!yellow, urlcolor=blue!95!black!85!yellow} 
\usepackage{physics}   
\usepackage{newtxtext} 
\usepackage[cmintegrals]{newtxmath} 
\usepackage{booktabs}  
\usepackage{multirow}  
\usepackage{xcolor}    
\usepackage{float}     

\usepackage{tikz}
\newcommand{\orcidicon}{%
    \begin{tikzpicture}
        \draw[lime, fill=lime] (0,0) circle [radius=0.16] node[white] {{\fontfamily{qag}\selectfont \tiny ID}};
        \draw[white, fill=white] (-0.0625,0.006) circle [radius=0.007];
    \end{tikzpicture}
    \hspace{-2.9mm}
}
\foreach \x in {A, B}{\expandafter\xdef\csname orcid\x\endcsname{\noexpand\href{https://orcid.org/\csname orcidauthor\x\endcsname}{\noexpand\orcidicon}}}

\newcommand{\iitm}{School of Mechanical and Materials Engineering, Indian Institute of Technology Mandi, Kamand 175075, India}

\newcommand{\lz}{LiZnAs}
\newcommand{\sa}{ScAgC}
\newcommand{\kph}{$\kappa_{\emph{ph}}$}
\newcommand{\eph}{$\emph{e-ph}$}
\newcommand{\cm}{cm$^{-3}$}
\newcommand{\cvs}{cm$^{2}$ V$^{-1}$ s$^{-1}$}

\newcommand{\mue}{$\mu_{\mathrm{e}}$}
\newcommand{\muh}{$\mu_{\mathrm{h}}$}

\preprint{APS/123-QED}

\begin{document}

\title{Significant first-principles electron--phonon coupling effects in the ${\rm LiZnAs}$\\ and ${\rm ScAgC}$ half-Heusler thermoelectrics}

\author{Vinod Kumar Solet \orcidA{}}
\email{vsolet5@gmail.com}
\author{Sudhir K. Pandey \orcidB{}}
\email{sudhir@iitmandi.ac.in}
\affiliation{\iitm}
\date{\today}

\begin{abstract}
Half-Heusler (hH) compounds are currently considered promising thermoelectric (TE) materials owing to their favorable thermopower and electrical conductivity. Accurate estimates of their TE performance are therefore highly desirable and require a detailed microscopic understanding of the mechanisms that limit carrier transport. To enable such estimations, we carry out comprehensive first-principles computations of the electron-phonon ($\emph{e-ph}$) interactions in two hH semiconductors (LiZnAs and ScAgC). Our study first investigates their electron and phonon dispersions and then examines the temperature-induced renormalization of the electronic states within the non-adiabatic Allen–Heine–Cardona formalism. We then solve the Boltzmann transport equation (BTE) both iteratively and within multiple relaxation-time approximations (RTAs) to evaluate the carrier transport properties. Phonon-limited electron and hole mobilities computed using the linearized self-energy and momentum RTAs (SERTA and MRTA) are compared with the iterative BTE (IBTE) results. The electrical transport coefficients for TE performance are subsequently evaluated under these two RTA schemes and compared with constant RTA (CRTA) results. The lattice thermal conductivity, determined from phonon-phonon interaction, is further reduced through nanostructuring. In bulk form, \lz\ (\sa) attains a maximum figure of merit $zT$ of $\sim$1.05 (0.78) at 900 K for an electron concentration of 10$^{18}$ (10$^{19}$) cm$^{-3}$ under the MRTA. For a 20-nm nanostructured sample, the corresponding $zT$ increase to $\sim$1.53 (1.0). The remarkably high $zT$ achieved through inherently present phonon-induced electron scattering effects, combined with grain-boundary engineering, opens a promising path for discovering highly efficient and accurate next-generation hH TEs.

\end{abstract}

\maketitle

\section{Introduction} 
\setlength{\parindent}{3em}
Half-Heusler (hH) semiconductors have received significant attention as promising thermoelectric (TE) materials for medium- and high-temperature clean energy-harvesting technology \cite{dong2022half,bos2014half,zhu2018discovery,zhu2015high,zhu2023half,fu2015band,fu2015realizing,yang2008evaluation,samsonidze2018accelerated,rogl2023development}. The efficiency of TE materials is given by the dimensionless figure of merit $zT$ = $\sigma$S$^2$T/$\kappa$, where $\sigma$ is the electrical conductivity, S is the Seebeck coefficient, T is the temperature, and $\kappa$ = ($\kappa_e$ + \kph) is the total thermal conductivity, comprising the electronic ($\kappa_e$) and lattice (\kph) contributions \cite{shastri2021theory}. The hH materials usually exhibit very promising electron transport properties, particularly high $\sigma$ and S values, as seen in compounds like ZrNiSn and cobalt-based hH alloys, which contribute to a high TE power factor $\sigma$S$^2$ \cite{yang2008evaluation,uher1999transport,shastri2020thermoelectric,zeeshan2017ab,zhu2017compromise,zhou2018large,ye2025superior}. Interestingly, hHs also exhibit a wide range of \kph, making these compounds attractive for various energy-related applications \cite{graf2011simple,uher1999transport,rogl2023development}. However, the inherently 2-4 times higher $\kappa$ values of hHs compared to other cutting-edge TEs \cite{uher1999transport} poses a major challenge to achieving high TE conversion efficiency \cite{carrete2014finding,casper2012half,xie2012recent}. While maximizing the $zT$ by tweaking electrical coefficients ($\sigma$, S, $\kappa_e$) is often complex and limited by their interrelation, and thus reducing \kph\ is considered a more effective route without significantly affecting the other factors \cite{beretta2019thermoelectrics,pecunia2023roadmap}. Several promising approaches such as grain boundary and doping engineering have been proposed over the years to lower $\kappa$ in hHs \cite{sakurada2005effect,liu2021synergistically,tranas2022attaining,joshi2011enhancement, zhu2017compromise}. At the same time, understanding the microscopic mechanisms for accurate estimation of electric transport coefficients is also highly desirable for predicting favorable TE materials. Therefore, the microscopic physics underlying both particles (electron and phonon) transport needs to be well understood \cite{park2025advances}, not only for fundamental scientific research but also for advancing practical TE device applications.

From the well-known relation $\sigma$ = $ne\mu$ \cite{ashcroft1976}, where $n$ is the carrier concentration, $e$ is the charge of carrier, and $\mu$ is the carrier mobility, a more accurate $zT$ effectively depends on the precise determination of $\mu$. In first principles density functional based calculations, obtaining transport properties such as $\mu$ and $\kappa$ is quite challenging, as it requires a detailed understanding of how particles are scattered by each other or by other particles \cite{bernardi2016first}. These scattering mechanisms can be understood from the Boltzmann transport theory \cite{zhou2016first,ponce2020first}. For mid- and high-T TE applications, the $\mu$ value is significantly affected by electron-phonon (\eph) scattering in semiconductors \cite{giustino2017electron,ziman1960}. \eph\ coupling also plays a crucial role for the T-dependent renormalization of band structure \cite{giustino2017electron,giustino2010electron,antonius2014many,ponce2015temperature,ponce2025verification}. First-principles \eph\ interaction (EPI) calculations demonstrate that electron lifetime varies significantly with electron energy and carrier concentration \cite{samsonidze2018accelerated,giustino2017electron}, suggesting the need to go beyond the commonly used constant relaxation time approximation (RTA) to effectively screen and optimize TE materials. Therefore, analyzing microscopic EPI mechanisms is favorable for accurately obtaining the electrical components of $zT$. On the other hand, phonon-phonon interaction (PPI) dominates in the thermal transport of semiconductors \cite{zhou2016first}. In TE materials, the electron mean-free-path (MFP) is much smaller than the phonon MFP \cite{qiu2015first,minnich2009bulk}. Therefore, the use of nanostructuring techniques with grain sizes above the electron MFP can be the most efficient way to reduce \kph\ \cite{solet2025band}. Thus, combining the effects of EPI on electrical transport and PPI on phonon transport can effectively aid in accurately identifying and designing hH TE materials.

Well-established publicly available software packages for obtaining electronic transport \cite{claes2025phonon}, each with different capabilities and strengths, include LanTRAP \cite{wang2018lantrap}, EPIC STAR \cite{deng2020epic}, ElecTra \cite{graziosi2023electra}, PERTURBO \cite{zhou2021perturbo}, EPW \cite{ponce2016epw} and TOSSPB \cite{pohls2022tosspb}. LanTRAP employs a Landauer mode-counting scheme without first-principles \eph\ scattering. EPIC STAR obtains energy-dependent relaxation times using density functional perturbation theory (DFPT) together with generalized Eliashberg function for short-range \eph\ scattering and analytical expressions for long-range \eph\ and electron–impurity scattering, and computes transport within the semi-classical Boltzmann transport equation (BTE). ElecTra solves the linearized BTE in the RTA using full-band density functional theory (DFT) inputs combined with model scattering mechanisms such as deformation-potential acoustic and optical phonons, polar optical phonons, ionized impurities, and alloy disorder. PERTURBO and EPW, in contrast, perform fully \emph{ab initio} transport by interpolating DFPT \eph\ matrix elements with Wannier functions and solving the BTE. TOSSPB is a simplified single-parabolic-band model aimed at scattering-dependent TE optimization. In comparison, our work includes all phonon-scattering channels directly from DFPT and solves the BTE both iteratively and within RTA entirely within ABINIT \cite{brunin2020phonon,gonze2020abinit}, providing a fully first-principles description of phonon-limited charge transport using Fourier interpolation. The approach implemented in ABINIT, as thoroughly described in Refs. \cite{brunin2020phonon,ponce2015temperature,ponce2014temperature,ponce2014verification}, achieves performance comparable to that of Wannier-based packages, without requiring the use of Wannier functions.

The hH materials, composed of one main-group element and two transition metals, offer a wide range of tunability, among which 18 valence electron count (VEC) systems have gained considerable interest for waste-heat recovery applications \cite{xie2012recent,gautier2015prediction,yang2008evaluation,berland2019thermoelectric}. Enhancing the hH TEs properties can achieved either by doping existing compounds \cite{casper2012half} or by identifying novel systems with superior properties. Systems with 8 VEC fit well into the latter category, as they have received relatively less attention in TEs compared to 18 VEC alloys \cite{ciftci2016electronic,barth2010investigation,vikram2019accelerated,toberer2009thermoelectric,yang2008evaluation}. Our literature survey indicates that most existing studies on 8 VEC alloys estimate the transport using constant RTA (CRTA) \cite{madsen2006automated,vikram2019accelerated,ciftci2016electronic,toberer2009thermoelectric,solet2022first,sahni2020reliable,yadav2015first,nazir2024tunable,azouaoui2022first,jolayemi2021investigation}. Previous studies on TEs show that the energy-dependent electron relaxation time within EPI significantly affects both $\sigma$ and $\kappa_e$ compared to CRTA-based values \cite{fiorentini2016thermoelectric,solet2025band}. Importantly, the S values are also influenced and agree better with experimental results within EPI \cite{solet2025band}. For 18 VEC compounds \cite{samsonidze2018accelerated,graziosi2020material}, similarly substantial differences between CRTA and energy-dependent scattering–based charge transport have been reported for both $n$-type and $p$-type samples. Several recent works in Refs.~\cite{witkoske2017thermoelectric,graziosi2019impact,graziosi2020material,rudderham2021ab,rudderham2020analysis} on different materials (including hHs) have also demonstrated that the CRTA can lead to less accurate predictions of TE transport. Two of these studies \cite{witkoske2017thermoelectric,graziosi2019impact} show that intervalley and interband \eph\ scattering, as well as the full energy and momentum dependence of the relaxation time, can significantly alter power factors and carrier-density trends. Model-based analyses in the other two works \cite{rudderham2021ab,rudderham2020analysis} further demonstrate that different simplified scattering assumptions (CRTA, constant mean free path, density-of-states-based scattering) can yield markedly different TE responses. These findings emphasize the need for fully \textit{ab initio} treatments of EPIs, as applied in this work.

To fulfill this research gap, we have studied the effect of \eph\ coupling on electronic band structure and electrical transport properties of two 8 VEC-based \lz\ and \sa\ hH materials using advanced \emph{ab initio} many-body perturbation theory (MBPT) calculations. The presence of complex valance bands (VBs) in these hH materials \cite{solet2025many} makes them an excellent platform for assessing the impact of scattering physics on their electronic and transport properties. The zero-point renormalization (ZPR) and T dependence band gaps up to 900 K using the EPI calculations within the Allen–Heine–Cardona (AHC) theory have been analyzed. A large ZPR correction to the DFT gap is observed in \sa\ than in \lz. We then employ the linearized BTE under the self-energy and momentum RTAs [SERTA and MRTA], as well as the iterative BTE (IBTE), within EPI to calculate the electron and hole mobilities (\mue\ and \muh). The \muh\ values are significantly lower than the \mue\ values in both hHs from SERTA, MRTA and IBTE, possibly can be due to presence of complex and highly degenerate bands around the topmost VB region. This feature is a clear indication of dominant $n$-type conduction over $p$-type. Furthermore, S, $\sigma$ and $\kappa_e$ values are significantly affected by the choice of RTAs such as CRTA, SERTA and MRTA. To estimate $zT$, \kph\ is calculated considering three-phonon scattering and is further reduced by incorporating grain-size effects through phonon-boundary scattering. Among the RTAs, the maximum $zT$ is found to be $\sim$1.05 (0.78) for an electron doping of 10$^{18}$ (10$^{19}$) cm$^{-3}$ under MRTA for \lz\ (\sa). These values further increase to $\sim$1.53 (1.0) when a grain size of 20 nm is considered. These results highlight the significant role of different electron and phonon scattering mechanisms in achieving more accurate and high $zT$ values (grater than 1) in these hHs. We hope this study provides valuable guidance for the discovery of novel, highly efficient 8 VEC hH materials.

\section{THEORY AND METHODS} \label{sec:thmet}

\subsection{Electron-phonon renormalization} \label{sec:epr}
The \eph\ self-energy due to EPI, $\Sigma^{\mathrm{ep}}_{n\textbf{k}}$, can be decomposed, based on the first order perturbation theory \cite{giustino2017electron}, into frequency dependent Fan-Migdal (FM) and the static and Hermitian Debye-Waller (DW) parts \cite{ponce2015temperature,ponce2014verification,antonius2015dynamical,giustino2010electron,cannuccia2011effect,ponce2014temperature},
\begin{eqnarray} \label{eq:se}
\Sigma^{\mathrm{ep}}_{nn'\textbf{k}}(\mathrm{\omega, T}) = \Sigma^{\mathrm{FM}}_{nn'\textbf{k}}(\mathrm{\omega, T}) +  \Sigma^{\mathrm{DW}}_{nn'\textbf{k}}(\mathrm{T}),                                
\end{eqnarray} 
\begin{align}
\Sigma_{nn'\mathbf{k}}^{\mathrm{FM}}(\omega,T) 
&= \frac{1}{N_q} \sum_{\substack{\mathbf{q}\nu m\\\kappa\alpha\kappa'\beta}} 
   \frac{1}{2\omega_{\mathbf{q}\nu}}\,g^{*}_{mn\kappa\alpha}(\mathbf{k},\mathbf{q})
   \,g_{mn'\kappa'\beta}(\mathbf{k},\mathbf{q}) \notag \\
&\quad\times e^{*}_{\kappa\alpha\nu}(\mathbf{q}) e_{\kappa'\beta\nu}(\mathbf{q}) \label{eq:fmse1} \\
&\quad \sum_{\pm}
\frac{n_{\mathbf{q}\nu}(T) +  \bigl[1 \pm (2f_{m\mathbf{k}+\mathbf{q}}(T) - 1 ) ] /2 }
     {\omega - \varepsilon_{m\mathbf{k}+\mathbf{q}} \pm \omega_{\mathbf{q}\nu} + i\eta}, \notag
\end{align}
\begin{equation}
\begin{aligned} \label{eq:dwse}
\Sigma_{nn'\mathbf{k}}^{\mathrm{DW}}(T) 
&= \frac{1}{N_q} \sum_{\mathbf{q}\nu} \frac{n_{\mathbf{q}\nu}(T) + \tfrac{1}{2}}{2\omega_{\mathbf{q}\nu}}\,D_{nn'\nu}(\mathbf{k},\mathbf{q}), 
\end{aligned}
\end{equation}
where $n$ and $n'$ refer bare electronic eigenstates, \textbf{k} is the electron momentum, $\omega_{\mathbf{q}\nu}$ denotes the phonon frequency for mode index $\nu$ and wavevector \textbf{q}, $f_{m\mathbf{k}+\mathbf{q}}$ ($n_{\mathbf{q}\nu}$) is the electron (phonon) occupation number at temperature T, $\eta$ is the positive real infinitesimal, $N_q$ is the total number of phonon momenta for discretizing the first Brillouin zone (BZ), $e_{\kappa\alpha\nu}$(\textbf{q}) is the displacement vector for atom $\kappa$ in the Cartesian direction $\alpha$ with mode $\nu$. 

To obtain the FM and the DW self-energies, one needs to estimate the following two types of matrix elements: 
\begin{equation}  \label{eq:gmn}
g_{mn\kappa\alpha}(\mathbf{k}, \mathbf{q})
= \langle u_{m\mathbf{k}+\mathbf{q}} \mid V_{\mathbf{q}\kappa\alpha} \mid u_{n\mathbf{k}} \rangle,
\end{equation}
and
\begin{equation}
\begin{aligned} \label{eq:dnn}
D_{nn'\nu}(\mathbf{k},\mathbf{q})
&\equiv \sum_{\kappa\alpha\kappa'\beta}
\langle u_{n\mathbf{k}} | \partial_{-\textbf{q}\kappa\alpha} \partial_{\textbf{q}\kappa'\beta} V^{KS} | u_{n'\mathbf{k}} \rangle\, \\
&\quad\times e^{*}_{\kappa\alpha\nu}(\mathbf{q})\, e_{\kappa'\beta\nu}(\mathbf{q}),
\end{aligned}
\end{equation}
where $u_{n\textbf{k}}$ is the periodic function of the Kohn-Sham (KS) Bloch wave function with the associated nonlocal part of the KS potential $V (\textbf{r}, \textbf{r}')$. Equation~\hyperref[eq:dnn]{\eqref{eq:dnn}} depends on the second order derivative of the KS potential $V^{KS}$ with respect to the ion displacements which are generally challenging to compute. To circumvent this difficulty, the matrix $D$ is evaluated within the rigid-ion approximation (RIA), where the second-order matrix elements are expressed in terms of the first-order ones \cite{ponce2015temperature,ponce2025verification}.  
\begin{equation} \label{eq:dnn1}
\begin{aligned}
D_{nn\nu}^{\mathrm{RIA}}(\mathbf{k},\mathbf{q})
&\equiv -\sum_{\substack{m\neq n\\\kappa\alpha\kappa'\beta}}
    g_{mn\kappa\alpha}^*(\mathbf{k},\Gamma)\,
    g_{mn\kappa'\beta}(\mathbf{k},\Gamma)\, \\
&\quad\times
    \frac{e_{\kappa\alpha\nu}^*(\mathbf{q}) e_{\kappa\beta\nu}(\mathbf{q})
          + e_{\kappa'\alpha\nu}^*(\mathbf{q}) e_{\kappa'\beta\nu}(\mathbf{q})}
         { \varepsilon_{n\textbf{k}} - \varepsilon_{m\textbf{k}} }.
\end{aligned}
\end{equation}

By using Eq.~\hyperref[eq:dnn1]{\eqref{eq:dnn1}} in Eq.~\hyperref[eq:dwse]{\eqref{eq:dwse}} for $n$ = $n'$, the diagonal elements of the DW term are obtained. The quasiparticle (QP) energy due to \eph\ coupling can then be estimated from the real part of Eq.~\hyperref[eq:se]{\eqref{eq:se}} at the bare band energy $\varepsilon_{n\textbf{k}}$, assuming non-diagonal terms are negligible. Here, the QP energy is obtained through two different approaches. In the first approach, the EPR energy at T using \emph{on-the-mass-shell} (OTMS) approximation is given by \cite{cannuccia2012zero}, 
\begin{eqnarray}\label{eq:otms}
\varepsilon^{OTMS}_{n\textbf{k}}(\mathrm{T}) =  \varepsilon_{n\textbf{k}} + \mathrm{Re}\Big[\Sigma_{nn\textbf{k}}^{\mathrm{ep}}(\varepsilon_{n\textbf{k}},\mathrm{T})\Big].           
\end{eqnarray}
Alternatively, if the QP energy is close to the bare energy, $\Sigma_{nn'\textbf{k}}^{\mathrm{ep}}(\omega,\mathrm{T})$ can be expended in a Taylor series around $\omega$ = $\varepsilon_{n\textbf{k}}$ and evaluated it at $\omega$ = $\varepsilon_{n\textbf{k}}$(T) to solve the linearized quasiparticle equation (LQE) as, 
\begin{align}
\varepsilon^{LQE}_{n\mathbf{k}}(T) &= \varepsilon_{n\mathbf{k}} + Z_{n\mathbf{k}}\,\mathrm{Re}\bigl[\Sigma^{\mathrm{ep}}_{nn\mathbf{k}}(\varepsilon_{n\mathbf{k}},T)\bigr], \label{eq:lqe} \\
Z_{n\mathbf{k}} & \equiv \left(1 - \mathrm{Re}\!\left[\left.\frac{\partial \Sigma^{\mathrm{ep}}_{nn\mathbf{k}}(\omega, T)}{\partial \omega}\right|_{\omega = \varepsilon_{n\mathbf{k}}}\right]\right)^{-1}. \label{eq:znk}
\end{align}
The ZPR of the excitation energy is calculated as the difference between the energy obtained from the equations above at T = 0 and the bare KS eigenvalue. Overall, the Eqs.~\hyperref[eq:fmse1]{\eqref{eq:fmse1}}, \hyperref[eq:dwse]{\eqref{eq:dwse}}, \hyperref[eq:dnn1]{\eqref{eq:dnn1}} and \hyperref[eq:otms]{\eqref{eq:otms}} form the non-adiabatic version of the AHC theory \cite{ponce2015temperature}.

\subsection{Charge lifetime and transport} \label{sec:charge}
In a semiconductor, the mobility is directly related to the steady-state electric current and the electric field \textbf{E} \cite{ziman1960}. It can be obtained by taking the derivative of current with respect to the electric filed, which can be written within the Boltzmann transport formalism as \cite{ziman1960},
\begin{eqnarray} \label{eq:mue}
\mu_{\mathrm{e}, \alpha\beta} = -\frac{\sum_{n\in \mathrm{CB}} \int d\textbf{k} \upsilon_{n\textbf{k}, \alpha} \partial_{E_{\beta}}f_{n\textbf{k}}} {\sum_{n\in \mathrm{CB}} \int d\textbf{k} f_{n\textbf{k}}^{0}}
\end{eqnarray}
The summations are considered for the conduction band (CB) states for electron mobility ($\mu_{\mathrm{e}, \alpha\beta}$) calculation, and $\partial_{E_\beta}$ refers to $\partial/\partial E_\beta$. A similar expression, where the summation is restricted to states in the VBs, gives the hole mobility (\muh). The band velocity $\upsilon_{n\mathbf{k}, \alpha}$ is defined by the electron eigenvalue $\varepsilon_{n\mathbf{k}}$ for state $\ket{n\mathbf{k}}$, with $\hslash^{-1} \partial\varepsilon_{n\mathbf{k}}/\partial k_{\alpha}$. The presence of an electric field causes a deviation from the equilibrium occupation number, i.e., the Fermi-Dirac distribution $f_{n\mathbf{k}}^{0}$ to a nonequilibrium distribution function $f_{n\mathbf{k}}$ \cite{ashcroft1976}. The indices $\alpha$ and $\beta$ run over the three Cartesian directions ($x$, $y$, and $z$). In order to solve Eq.~\hyperref[eq:mue]{\eqref{eq:mue}}, we need to obtain $\partial_{E_{\beta}}f_{n\mathbf{k}}$, which is the linear response of $f_{n\mathbf{k}}$ to the \textbf{E} and can be derived starting from the electron BTE as \cite{ziman1960},  
\begin{align} \label{eq:bte}
(-e)\mathbf{E} \cdot \frac{1}{\hslash} \frac{\partial f_{n\mathbf{k}}}{\partial \mathbf{k}} 
= \frac{2\pi}{\hslash} \sum_{m\nu} \int \frac{d\mathbf{q}}{\Omega_{\text{BZ}}} |g_{mn\nu}(\mathbf{k}, \mathbf{q})|^2 \notag \\
\quad \times \Big\{ (1 - f_{n\mathbf{k}}) f_{m\mathbf{k}+\mathbf{q}} 
\delta(\varepsilon_{n\mathbf{k}} - \varepsilon_{m\mathbf{k}+\mathbf{q}} + \hslash \omega_{\mathbf{q}\nu})(1 + n_{\mathbf{q}\nu}) \notag \\
\quad + (1 - f_{n\mathbf{k}}) f_{m\mathbf{k}+\mathbf{q}} 
\delta(\varepsilon_{n\mathbf{k}} - \varepsilon_{m\mathbf{k}+\mathbf{q}} - \hslash \omega_{\mathbf{q}\nu}) n_{\mathbf{q}\nu} \notag \\
\quad - f_{n\mathbf{k}} (1 - f_{m\mathbf{k}+\mathbf{q}}) 
\delta(\varepsilon_{n\mathbf{k}} - \varepsilon_{m\mathbf{k}+\mathbf{q}} - \hslash \omega_{\mathbf{q}\nu})(1 + n_{\mathbf{q}\nu}) \notag \\
\quad - f_{n\mathbf{k}} (1 - f_{m\mathbf{k}+\mathbf{q}}) 
\delta(\varepsilon_{n\mathbf{k}} - \varepsilon_{m\mathbf{k}+\mathbf{q}} + \hslash \omega_{\mathbf{q}\nu}) n_{\mathbf{q}\nu} \Big\}.
\end{align}
The left term of Eq.~\hyperref[eq:bte]{\eqref{eq:bte}} denotes the non-collision term of BTE in a uniform and constant electric field, i.e., in the absence of magnetic field and temperature gradients. On the other hand, the right side represents the change in the distribution function due to \eph\ scattering into and out of the state $\ket{n\mathbf{k}}$, via emission or absorption of phonons. Finally, by taking the derivatives of Eq.~\hyperref[eq:bte]{\eqref{eq:bte}} with respect to \textbf{E}, the explicit expression of $\partial_{E_{\beta}}f_{n\mathbf{k}}$ can be estimated,
\begin{align} \label{eq:ibte}
\partial_{E_{\beta}} f_{n\mathbf{k}} 
= e \frac{\partial f_{n\mathbf{k}}^{0}}{\partial \varepsilon_{n\mathbf{k}}} \upsilon_{n\mathbf{k}, \beta} \tau_{n\mathbf{k}}^{0} 
+ \frac{2\pi}{\hslash} \tau_{n\mathbf{k}}^{0} \sum_{m\nu} \int \frac{d\mathbf{q}}{\Omega_{\text{BZ}}} |g_{mn\nu}(\mathbf{k}, \mathbf{q})|^2 \notag \\
\quad \times \left[ (1 - f_{n\mathbf{k}}^{0} + n_{\mathbf{q}\nu}) 
\delta(\varepsilon_{n\mathbf{k}} - \varepsilon_{m\mathbf{k}+\mathbf{q}} + \hslash \omega_{\mathbf{q}\nu})\qquad\qquad\quad\quad\quad \right. \notag \\
\quad \quad + \left. (f_{n\mathbf{k}}^{0} + n_{\mathbf{q}\nu}) 
\delta(\varepsilon_{n\mathbf{k}} - \varepsilon_{m\mathbf{k}+\mathbf{q}} - \hslash \omega_{\mathbf{q}\nu}) \right] 
\partial_{E_{\beta}} f_{m\mathbf{k+q}} \quad\quad\quad
\end{align} 
where the relaxation time is defined as,
\begin{equation} \label{eq:tau}
\begin{aligned}
\frac{1}{\tau_{n\mathbf{k}}^{0}} &= \frac{2\pi}{\hslash} \sum_{m\nu} \int \frac{d\mathbf{q}}{\Omega_{\text{BZ}}} |g_{mn\nu}(\mathbf{k}, \mathbf{q})|^2 \\
&\quad \times \left[ (1 - f_{m\mathbf{k+q}}^{0} + n_{\mathbf{q}\nu}) \delta(\varepsilon_{n\mathbf{k}} - \varepsilon_{m\mathbf{k}+\mathbf{q}} - \hslash \omega_{\mathbf{q}\nu}) \right. \\
&\qquad + \left. (f_{m\mathbf{k+q}}^{0} + n_{\mathbf{q}\nu}) \delta(\varepsilon_{n\mathbf{k}} - \varepsilon_{m\mathbf{k}+\mathbf{q}} + \hslash \omega_{\mathbf{q}\nu}) \right].
\end{aligned}
\end{equation}
Equation~\hyperref[eq:ibte]{\eqref{eq:ibte}} is the linearized BTE, which needs to be solved self-consistently for $\partial_{E_{\beta}}f_{n\mathbf{k}}$, and is commonly referred to as the IBTE. A simple approach can be considered in which only the first term on the right-hand side of Eq.~\hyperref[eq:ibte]{\eqref{eq:ibte}} is used, meaning the sum term is neglected. With this consideration, one can estimate the variation $\partial_{E_{\beta}}f_{n\mathbf{k}}$ without solving it iteratively. In this approximation, the relaxation time $\tau_{n\mathbf{k}}^{0}$ is directly associated with the imaginary part of the FM electron self-energy using Eq.~\hyperref[eq:fmse1]{\eqref{eq:fmse1}} via $(\tau_{n\mathbf{k}}^{0})^{-1}$ = 2Im$\Sigma_{nn\mathbf{k}}^{\mathrm{FM}}$ \cite{brunin2020phonon,solet2025band}, which gives the lifetime of the charged QP excitations due to EPI. The approximation involving the omission of the integral in Eq.~\hyperref[eq:ibte]{\eqref{eq:ibte}} is thus referred to as the SERTA. The mobility in Eq.~\hyperref[eq:mue]{\eqref{eq:mue}} can therefore be further updated as,
\begin{eqnarray} \label{eq:serta}
\mu_{\mathrm{e}, \alpha\beta} = -\frac{e}{n_{\mathrm{e}}\Omega} \sum_{n\in \mathrm{CB}}\int \frac{d\mathbf{k}}{\Omega_{\text{BZ}}} \frac{\partial f_{n\mathbf{k}}^{0}}{\partial \varepsilon_{n\mathbf{k}}} \upsilon_{n\textbf{k}, \alpha} \upsilon_{n\textbf{k}, \beta} \tau_{n\mathbf{k}}^{0},
\end{eqnarray}
where $n_{\mathrm{e}}$ denotes the electron density, and $\Omega$ and $\Omega_{\text{BZ}}$ represent the volumes of the crystalline unit cell and first BZ, respectively. In the SERTA, the BTE is solved in a simplified, non-iterative form \cite{ponce2020first}. It is also equivalent for only accounting out-of-state transitions \cite{ponce2020first}. However, to also capture the effects of back-scattering, the MRTA is introduced. It modifies the scattering probability through a geometrical factor that weights forward-scattering more heavily. In this process, the integrand in Eq.~\hyperref[eq:tau]{\eqref{eq:tau}} is now weighted by the factor \cite{ponce2020first,li2015electrical,claes2022assessing}, 
\begin{equation}\label{eq:almrt}
\alpha^{\mathrm{MRTA}}_{mn}(\mathbf{k},\mathbf{q})
= \left( 1 - \frac{\mathbf{v}_{n\mathbf{k}}\!\cdot\!\mathbf{v}_{m\mathbf{k}+\mathbf{q}}}
{|\mathbf{v}_{n\mathbf{k}}|^{2}} \right).
\end{equation}


Once the electron energy dependent lifetime $\tau_{n\mathbf{k}}^{0}$ is obtained, all three electrical components of TE materials such as S, $\sigma$, and $\kappa_e$ are easily estimated within the electron BTE \cite{solet2025band}. 

\subsection{Phonon thermal conductivity} \label{sec:kph}
Using single-mode relaxation-time (SMRT), $\tau_{\lambda}$, \kph\ tensor can be easily written by solving phonon linearized BTE \cite{togo2015distributions},
\begin{eqnarray} \label{eq:kph}
  \kappa_{ph} = \frac{1}{NV_{0}}\sum_{\lambda}C_{\lambda}\textbf{v}_{\lambda}\otimes \textbf{v}_{\lambda}\tau_{\lambda},
\end{eqnarray}
where $N$ and $V_{0}$ are the number of unit cells and volume of unit cell, respectively. Here $\lambda$ is the phonon mode, denoted by the \textbf{q}$\nu$. The mode-dependent specific heat $C_{\lambda}$ and phonon group velocity  $\textbf{v}_{\lambda}$ can be estimated directly from the solution of eigenvalue problem for phonons \cite{togo2015distributions,solet2023ab}. The SMRT can be obtained using imaginary part of phonon self-energy, $\Gamma_{\lambda}(\omega = \omega_{\lambda})$, under MBPT via Fermi's golden rule as  \cite{togo2015distributions},
\begin{eqnarray} \label{eq:phtau}
\tau_{\lambda}^{-1} = 2 \Gamma_{\lambda}(\omega =\omega_{\lambda})=\frac{36\pi}{\hslash^{2}}\sum_{\lambda^{'}\lambda^{''}} \Big| \Phi_{-\lambda\lambda^{'}\lambda^{''}}\Big|^{2}\biggl\{\bigl(n_{\lambda^{'}}+ n_{\lambda^{''}}+ 1\bigl) \nonumber \\ \times \delta\bigl(\omega - \omega_{\lambda^{'}} - \omega_{\lambda^{''}}\bigl) + \bigl(n_{\lambda^{'}} - n_{\lambda^{''}}\bigl) \quad\quad\quad\nonumber \\\times \Bigl[\delta\bigl(\omega + \omega_{\lambda^{'}} - \omega_{\lambda^{''}}\bigl) - \delta\bigl(\omega - \omega_{\lambda^{'}} + \omega_{\lambda^{''}}\bigl)\Bigl] \biggl\}, \quad \quad
\end{eqnarray}
in which $\Phi_{-\lambda\lambda^{'}\lambda^{''}}$ represents the interaction strength between phonon modes $\lambda$, $\lambda^{'}$ and $\lambda^{''}$.

\subsection{Computational details} \label{sec:comput}
The starting point for our first-principles MBPT-based EPI calculations is the DFT and DFPT, which provide the energies and wave functions of both particles, as well as the perturbation potentials used to compute Eq.~\hyperref[eq:gmn]{\eqref{eq:gmn}} \cite{gonze1997dynamical,baroni2001phonons,brunin2020phonon,ponce2015temperature}. We perturb all atoms in the primitive unit cell along all three directions to compute the force constants using DFPT. All \emph{ab initio} based EPI calculations are carried out using the Perdew-Burke-Ernzerhof (PBE) functional \cite{perdew1996generalized} in pseudo-potentials based Abinit software \cite{gonze2002first,gonze2020abinit}, with a kinetic energy cutoff of 40 Hartree applied for the truncation of plane-wave basis set. A $\Gamma$ centered 16 $\times$ 16 $\times$ 16 \textbf{k}-mesh and 8 $\times$ 8 $\times$ 8 \textbf{q}-mesh are used to obtain the ground-state electronic density and to perform DFPT phonon computations, respectively. 

For ZPR and T-dependent gap calculations, the scattering potentials are Fourier interpolated onto a dense 48 $\times$ 48 $\times$ 48 sampling in \textbf{q}-space, while the finer Bloch states are obtained on a similar grid in \textbf{k}-space via a non-self consistent calculation. Furthermore, accurate mobility calculations in Eq.~\hyperref[eq:serta]{\eqref{eq:serta}} must involve the large sampling density of \textbf{k} and \textbf{q} \cite{solet2025band}. Thus, the computed \eph\ scattering potentials on 8 $\times$ 8 $\times$ 8 \textbf{q}-mesh are Fourier interpolated onto 144 $\times$ 144 $\times$ 144 \textbf{q} grid mesh. Similarly, the large Bloch states are estimated through non-self consistent field calculations at 144 $\times$ 144 $\times$ 144 \textbf{k} points grid. The KS states used in the \eph\ self-energy calculations for solving Eq.~\hyperref[eq:tau]{\eqref{eq:tau}} are selected up to 0.7 (0.1) eV and 0.4 (0.08) eV above (below) the CB (VB) minimum (maximum), CBM (VBM), when \mue\ (\muh) is computed, which ensures good convergence in \lz\ and \sa, respectively. A small electron (hole) doping of $10^{15}$~\cm\ in the CB (VB) region is used to obtain the intrinsic mobility.

The \kph\ is estimated using the Phono3py program within finite displacement supercell method \cite{togo2015distributions}. A 2 $\times$ 2 $\times$ 2 (96 atoms) supercell (including interactions only up to third-nearest neighbors for anharmonic third-order force constants) is constructed and the resulting forces on these supercells are calculated using the Abinit software. A plane wave energy cutoff of 25 Hartree, a 4 $\times$ 4 $\times$ 4 \textbf{k}-point mesh, and a force convergence criteria of 5 $\times$ $10^{-8}$ Hartree/Bohr are employed in these force calculations. Furthermore, using the second- and third-order force constants, the BZ for phonon wave vectors is integrated on a well converged \textbf{q}-point grid of 23 $\times$ 23 $\times$ 23 to solve the \kph\ in Eq.~\hyperref[eq:kph]{\eqref{eq:kph}}. To obtain the phonon-scattering effect of grain boundaries, phonon-relaxation time is obtained by using Matthiessen's rule including boundary scattering rate $\upsilon_g/d$, where $\upsilon_g$ and $d$ are the phonon group velocity and grain size, respectively.

\begin{figure*}
\includegraphics[width=15.8cm, height=8.2cm]{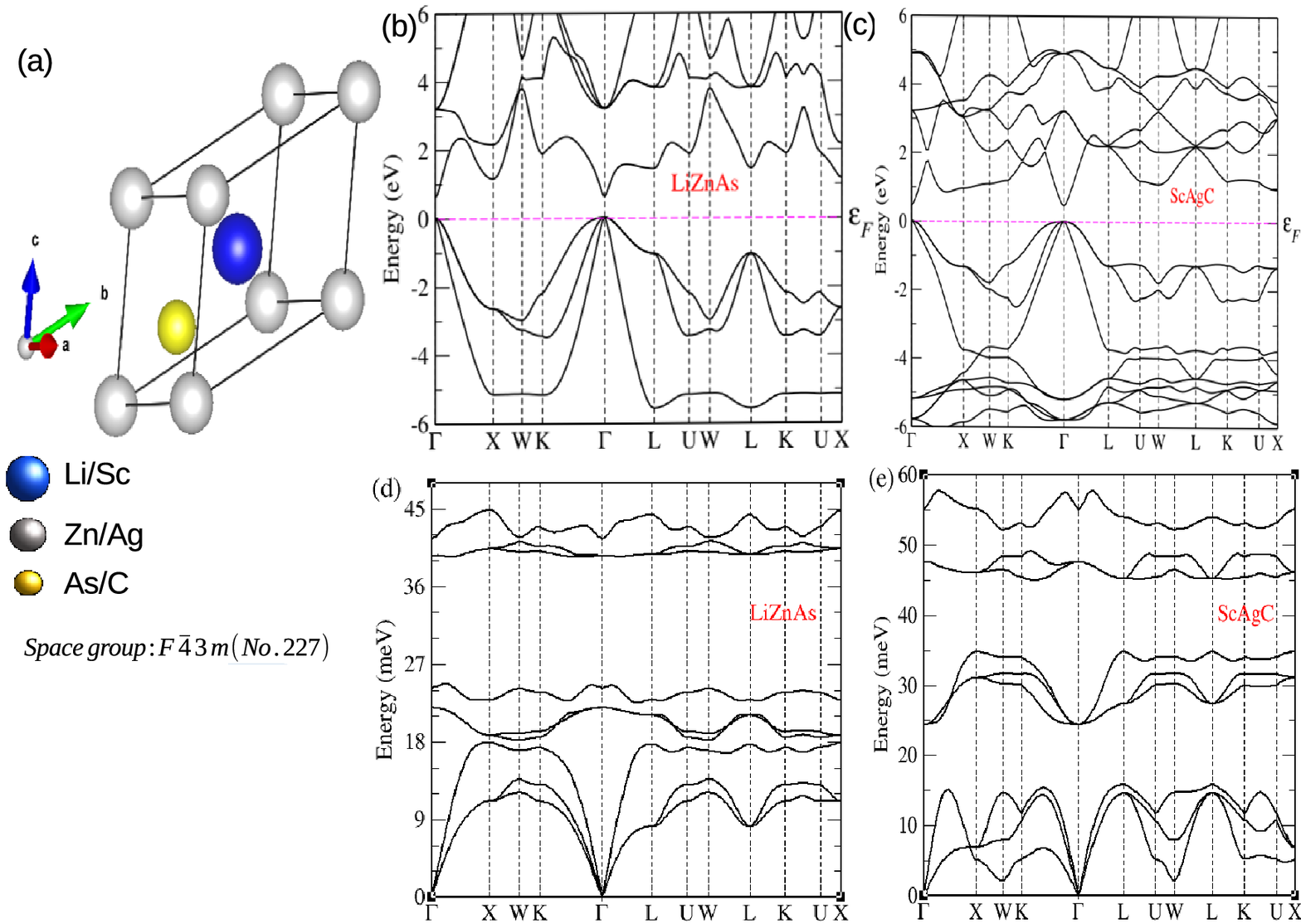} 
\caption{(a) Primitive unit cell used for the DFT, DFPT and EPI calculations. Blue, gray, and yellow symbols denote Li (Sc), Zn (Ag), and As (C) atoms, respectively, for \lz\ (\sa). Electronic band structures for \lz\ and \sa\ are shown in (b) and (c), respectively, and the corresponding phonon dispersions are presented in (d) and (e). The Fermi level ($\varepsilon_F$) is set at the top of the valence band. }
\label{fig:epd}
\end{figure*}

\section{Results and discussion} \label{sec:result}
\subsection{\label{sec:3a}Atomic structure, dispersion relations and EPI-induced renormalization}
The electron and vibrational dispersions are the primary quantities needed to understand the \eph\ scattering mechanisms in TE materials. All are carried out within the primitive unit cells. The atomic arrangements within the unit cells of both hHs are shown in Fig.~\hyperref[fig:epd]{\ref{fig:epd}(a)}, where both are found in the space group F$\bar{4}$3m (number 216) of the face-centered cubic (fcc) structure. This structure contains a basis of three atoms in the primitive cell, in which two atoms are transition or rare-earth metals. The Wyckoff positions of the atoms in \lz\ (\sa) are as follows: Li (Sc) at 4b (0.5, 0.5, 0.5), Zn (Ag) at 4a (0.0, 0.0, 0.0), and As (C) at 4c (0.25, 0.25, 0.25) \cite{kuriyama1994optical,solet2022first}. The \kph\ calculation is performed within a supercell constructed from the conventional cell of this primitive cell. The chosen lattice parameters are 5.94 \AA\ for \lz\ \cite{kuriyama1994optical} and 5.59 \AA\ for \sa\ \cite{solet2022first}. 

From the electronic dispersion shown in Figs.~\hyperref[fig:epd]{\ref{fig:epd}(b)} and \hyperref[fig:epd]{\ref{fig:epd}(c)}, both hHs exhibit a direct band gap semiconducting nature, with gap values of $\sim$0.6 eV for \lz\ and $\sim$0.46 eV for \sa\ at the $\Gamma$-point. These values are consistent with previous studies \cite{solet2025many,solet2022first}. The triply degenerate VBs at $\Gamma$ are present in both systems and become non-degenerate when moving away from this high-symmetry point. A single parabolic band near the CBM can also be observed in both materials. This indicates that near the Fermi level ($\varepsilon_F$), the CB is relatively simple and can be well described by a parabolic dispersion, whereas the VBs are more complex and often degenerate. Consequently, hole transport is typically subject to stronger scattering than electron transport. A key trait common to both hHs is that they contain several conduction valleys and flat dispersion, and their higher-lying bands are sufficiently energetic that they make little to no contribution to electrical conduction. These kind of CB and VB features inherently support both intravalley and intervalley scattering \cite{zhou2016ab,zhou2025electron,ding2021thermoelectric}. While such features can in principle provide additional scattering pathways, their relevance depends critically on the valley-to-valley energy separation \cite{zhou2016ab}. Nevertheless, the presence of the higher-lying valleys can become important at elevated temperatures or high doping levels if the $\varepsilon_F$ shifts sufficiently upward. Overall, the band features near $\varepsilon_F$ around the $\Gamma$-point; particularly the light-mass, parabolic CBM and the heavier, degenerate VBs; play a crucial role in determining carrier effective masses and the dominant \eph\ scattering mechanisms that govern various temperature-dependent phenomena, which are discussed in later sections.

The phonon dispersion is further examined to assess the dynamical stability of the two compounds and to elucidate their \eph\ scattering mechanisms [see Figs.~\hyperref[fig:epd]{\ref{fig:epd}(d)} and \hyperref[fig:epd]{\ref{fig:epd}(e)}]. Because each primitive cell contains three atoms, nine phonon branches (three acoustic and six optical) emerge. The absence of imaginary frequencies throughout the BZ confirms that both \lz\ and \sa\ are dynamically stable in the fcc phase. In our phonon calculations, due to the polar nature of the materials, the non-analytical term correction is included to account for the effect of the polarization field on the optical phonon branches around the $\Gamma$-point. The maximum phonon energies are found to be $\sim$45 meV in \lz\ and $\sim$58 meV in \sa. The zone-edge phonons in \lz, particularly those near the $X$ point that mediate intervalley scattering, possess energies beginning at $\sim$11 meV, which is higher than those typically found in \sa. In \lz, phonon branches near 39 meV exhibit weak dispersion, whereas in \sa\ a similarly flat region occurs near 45 meV. Such flat features enhance the phonon density of states and are expected to increase the corresponding phonon-mediated scattering rates. A notable feature in the phonon dispersions is the emergence of energy gaps, which originate from the large mass differences between the constituent atoms. Furthermore, the energy gap between the highest acoustic and lowest optical branches is smaller in \lz\ than in \sa, which could be important for understanding possible scattering mechanisms such as PPI.

Since band gap information is crucial for most T-dependent transport phenomena, understanding the renormalization of the electronic band gap is essential for high-T TE applications. In this work, T dependence of the band gap is obtained using the OTMS and LQE approaches, both of which are formulated within the non-adiabatic AHC formalism \cite{ponce2015temperature}. Based on the previous studies \cite{ponce2015temperature,miglio2020predominance}, only the non-adiabatic version of the AHC equations can be reliably used when computing the ZPR for infrared-active (IR-active) materials. In our study, the calculated non-zero Born effective charges for atoms in both \lz\ and \sa\ hH materials indicate that these compounds exhibit IR activity. This justifies the use of the non-adiabatic ZPR formulation in our calculations. 

\begin{figure}[ht]
\includegraphics[width=8.5cm, height=7.0cm]{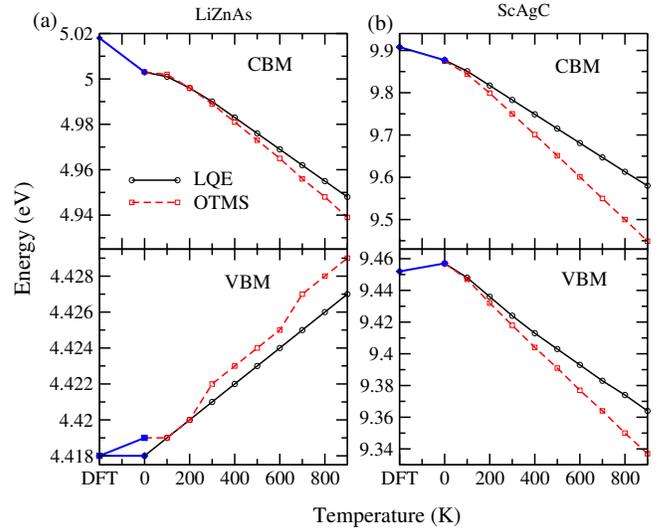} 
\caption{Renormalization and temperature dependence of the VBM and CBM states at the $\Gamma$ point obtained from both LQE and OTMS methods for (a) \lz\ and (b) \sa. The blue lines indicate the ZPR, connecting the DFT eigenvalues with the renormalized energy at 0 K. }
\label{fig:disp}
\end{figure}

It is observed from Table~\ref{tab:gap} for both approaches that the band gap decreases with increasing T, which can be explained by its relationship with T as described by the Varshni equation \cite{giustino2017electron}. Due to zero-point lattice vibrations, DFT gaps are reduced to $\sim$0.585 eV for \lz and 0.417-0.42 eV for \sa. The ZPR correction to the DFT band gap is found to be around 15-16 meV for \lz\ and 36-39 meV for \sa. As shown in Fig.~\hyperref[fig:disp]{\ref{fig:disp}}, this correction in \lz\ arises entirely from a downward shift of the CBM in both approaches, whereas a $\sim$31–33 meV shift of the CBM is observed for \sa. These large CBM correction in hHs can be understood from their band structures. The CB states in \sa\ are noticeably flatter than those in \lz, which can increase the density of available final states and could enhance the \eph\ self-energy. in addition, a larger spectral-weight transfer is observed at higher T, possibly due to increased lattice vibration strength and enhanced \eph\ scattering. At a given T, however, the magnitude of this weight transfer is smaller in \lz\ than in \sa, indicating there might be comparatively weaker \eph\ scattering at the $\Gamma$-point in \lz. It is also important to note that the ZPRs in Table~\ref{tab:gap} are typically smaller than those reported for other classic semiconductors such as Si, SiC, MgO, and diamond~\cite{zhang2020temperature}. This could be because the zero-point motion effect has a greater influence on the band gap in semiconductors composed of light atoms, owing to their larger atomic displacements. In contrast, the relatively heavy atoms present in our hH materials lead to smaller atomic displacements and weaker \eph\ coupling, resulting in smaller ZPR values. We also note that the present AHC calculations are performed within the RIA and neglect the non-diagonal components of the DW term. While this simplification is commonly adopted, it may become less accurate in materials where the band-edge states are closely spaced or exhibit strong interband coupling. In particular, the denser manifold of states near the VBM in both materials suggests that the omitted non-diagonal DW contributions may still play a minor role \cite{gonze2011theoretical}. Therefore, the ZPR values reported here should be interpreted with this limitation in mind. However, ZPR is well described in comparison with experimental results for various covalent materials when using non-adiabatic AHC calculations \cite{miglio2020predominance}.

\begin{table}[ht]
\caption{DFT band gap (without EPR), and ZPR correction along with temperature-dependent variations of the band gap obtained using the linearized quasiparticle equation (LQE) and on-the-mass-shell (OTMS) methods. All values are in eV.}
\centering
\setlength{\tabcolsep}{4.0pt} 
\renewcommand{\arraystretch}{1.1} 

\begin{minipage}{0.5\textwidth} 
\begin{tabular}{@{\extracolsep{\fill}} c c c c c c c c c c c c }
\hline\hline 
Temperature && \multicolumn{3}{c}{\lz} &  & \multicolumn{4}{c}{\sa} &  & \\
\cline{3-5} \cline{7-10}
(in K) && LQE && OTMS && LQE &&& OTMS && \\
\hline   
ZPR          && $-$0.015 && $-$0.016 && $-$0.036 &&& $-$0.039 && \\
Without EPR  &&  0.600 &&  0.600 &&  0.456 &&&  0.456 && \\
0            &&  0.585 &&  0.584 &&  0.420 &&&  0.417 && \\
100          &&  0.583 &&  0.582 &&  0.403 &&&  0.396 && \\
200          &&  0.577 &&  0.576 &&  0.381 &&&  0.366 && \\
300          &&  0.569 &&  0.567 &&  0.358 &&&  0.332 && \\
400          &&  0.561 &&  0.558 &&  0.335 &&&  0.297 && \\
500          &&  0.553 &&  0.549 &&  0.312 &&&  0.260 && \\
600          &&  0.545 &&  0.539 &&  0.288 &&&  0.224 && \\
700          &&  0.537 &&  0.529 &&  0.264 &&&  0.187 && \\
800          &&  0.529 &&  0.519 &&  0.239 &&&  0.150 && \\
900          &&  0.521 &&  0.510 &&  0.216 &&&  0.112 && \\
\hline\hline
\end{tabular}
\end{minipage}
\label{tab:gap}
\end{table}
 
\begin{figure}[ht]
\includegraphics[width=8.3cm, height=9.5cm]{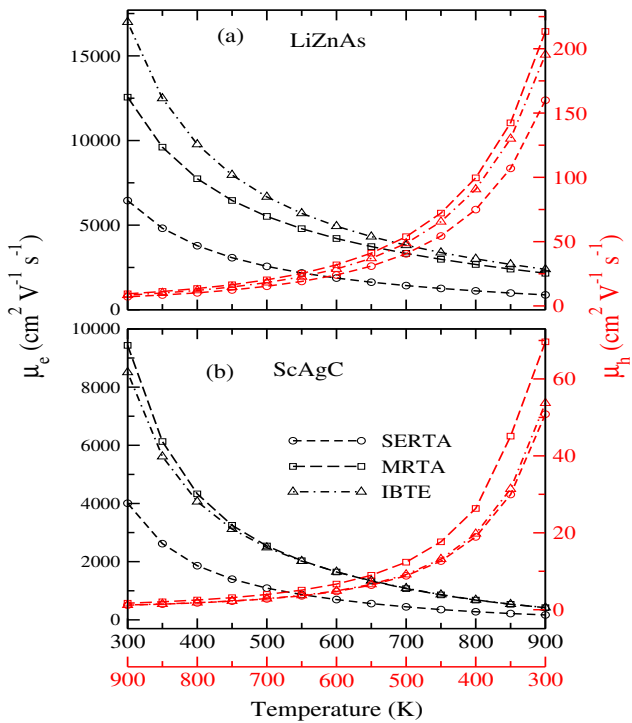} 
\caption{Temperature-dependent electron and hole mobilities using methods based on linearized (SERTA, MRTA) and iterative Boltzmann transport equation (IBTE) for (a) \lz\ and (b) \sa.} 
\label{fig:mob}
\end{figure}

\subsection{\label{sec:3b}Charge carrier mobility}

The mobility of charge carriers is one of the most important quantities for any semiconducting material, determining its suitability for applications in a wide variety of electronic and optoelectronic devices. Therefore, accurately determining the mobility is a key task, which requires extensive sampling of electron and phonon wave vectors in BZ. In the present work, only phonon-induced electron transitions are considered, and the effect of polaron formation is not included. In addition, electron–electron scattering and the phonon drag effect are also ignored. It is also important to note the the ZPR of both hH materials is found in the range of 0.01-0.04 eV. These values are less than or equal to the ZPR reported for GaAs (0.04 eV) \cite{grilli1992high,olguin2002electron}, which shares similarities in terms of band structure (and lattice parameters) with both hH compounds. Furthermore, previous studies on GaAs \cite{sjakste2015wannier} and TiO$_2$ \cite{verdi2015frohlich} have shown that, even without considering EPR due to EPI, the computed electron scattering rates remain reasonable, and thus our omission of EPR in the current calculations should still yield a reliable description of electrical transport in these hH materials. Consequently, phonon-limited mobility (and TE transport in Sec.~\ref{sec:3c}) are not expected to change significantly, as the other effects discussed here play only minor roles in electron transport for nondegenerate semiconductors at room temperature or above.  

The results obtained for \mue\ and \muh\ at electron concentration (n$_e$) of $10^{15}$~\cm\ using very fine sampling in both hH compounds are presented in Fig.~\hyperref[fig:mob]{\ref{fig:mob}}. It is important to note in both materials that the \mue\ is significantly higher than the \muh. This can be explained by the presence of more scattering channels for holes due to multiple VBs near the VBM, while fewer scattering processes exist around the CBM for electron transport because of the single CB. This results in a shorter $\tau_{n\mathbf{k}}^{0}$ for holes than for electrons. Furthermore, both mobilities decrease with increasing T due to enhanced phonon scattering at higher T. At room-temperature, the obtained \mue\ values for \lz\ are $\sim$6500, 12560, and 17000 \cvs, as calculated using SERTA, MRTA and IBTE, respectively. These respective methods estimate \muh\ values as $\sim$160, 213, and 195 \cvs. This \mue\ result compares well with reported \mue\ values for GaAs (a similar compound in terms of band structure), which range from 7000 to 12000 \cvs\ \cite{brunin2020phonon}. Furthermore, the room-temperature \mue\ (\muh) values for \sa\ are determined to be $\sim$4010 (50), 9440 (70), and 8510 (54) from respective methods. It is observed that MRTA mobility is consistently higher than that predicted by the SERTA. This behavior is expected for materials in which a single band exists around $\Gamma$ (the CB in our case), where intravalley scattering largely dominates--particularly for small wave vectors $\mathbf{q}$, because the effective masses are lower than 1~\cite{brunin2020phonon,solet2022first}. When the relevant $\mathbf{q}$-vectors are small, the factor appearing in Eq.~\hyperref[eq:almrt]{\eqref{eq:almrt}} lies between 0 and 1, leading to a mobility enhancement when going from SERTA to MRTA. However, when compared with the IBTE, no general trend exists, as also observed in Fig.~\hyperref[fig:mob]{\ref{fig:mob}}. The MRTA may either underestimate or overestimate the mobility, indicating that the factor in Eq.~\hyperref[eq:almrt]{\eqref{eq:almrt}} can underestimate or overestimate the back-scattering processes. 

The region around the VBM in both hHs is predominantly \textit{p}-like, whereas the CBM has mainly \textit{s}-like character~\cite{solet2025many}. The much lower \muh\ observed here may be due to strong \eph\ coupling associated with the localized \textit{p} state \cite{atta2004electron}. In addition, the VBs are more anisotropic and strongly $\mathbf{k}$-dependent, with significant mixing of orbital characters across the BZ (as shown in our previous study~\cite{solet2025many}). Such anisotropy can enhance the variation in \eph\ matrix elements and opens additional scattering pathways, leading to higher scattering rates for holes compared with the relatively simple and isotropic CB. To further clarify this behavior, we analyze the temperature dependence of the mobility, fitted using $\mu \propto T^{n}$ for both \mue\ and \muh\ in the SERTA, MRTA, and IBTE plots shown in Fig.~\hyperref[fig:mob]{\ref{fig:mob}}. The fitted $n$ values for electrons range from $-1.1$ to $-1.8$ for \lz\ and $-2.5$ to $-2.6$ for \sa. The much less negative values in \lz\ indicate weaker \eph\ scattering compared to \sa, consistent with the behavior observed in the ZPR analysis in Sec.~\ref{sec:3a}. For holes, the fitted exponents lie between $n=-2.7$ and $-2.75$ in \lz\ and between $n=-3.3$ and $-3.46$ in \sa, confirming that hole transport experiences significantly stronger \eph\ scattering in both hHs, particularly in \sa. 
At higher T in \lz, all three methods provide almost similar \muh\ values, which means second term of Eq.~\hyperref[eq:ibte]{\eqref{eq:ibte}} has a negligible contribution as the T increases, and the forward scattering also becomes nearly equal to the backward scattering around VBM. In \sa, the MRTA and IBTE solutions provide almost identical \mue\ values, especially beyond 400 K, that means velocity factor compensates the effect of the second term of Eq.~\hyperref[eq:ibte]{\eqref{eq:ibte}} around CBM. While result of \muh\ observes the completely opposite behavior where the SERTA and IBTE methods give the almost same values, indicating that the hole-like region is not significantly influenced by the second term of Eq.~\hyperref[eq:ibte]{\eqref{eq:ibte}}. These results indicate that electron transport significantly outperforms hole transport, suggesting that n-type samples of both materials can exhibit superior TE performance. We note that the calculated mobilities represent intrinsic phonon-limited values, and additional scattering mechanisms such as impurities or grain boundaries especially at high doping levels would further reduce the mobility in real samples.

\subsection{\label{sec:3c}Thermoelectric performance parameters}

TE transport properties have been obtained using Boltzmann transport theory within various computational approaches. The most commonly used approximation in this theory for calculating electrical transport is CRTA, which assumes an energy-independent electron lifetime \cite{ziman1960}. However, in semiconductors, due to the strong dependence of the electron lifetime on the electronic eigenvalues, CRTA can be an inadequate approach \cite{ziman1960}. This inadequacy is often reflected in the significant discrepancies observed between CRTA-based results and experimental data \cite{solet2025band,nam2022low}. Therefore, it is crucial to account for intrinsic scattering phenomena to achieve accurate transport predictions, and in this work, a more demanding first principles \eph\ coupling is considered \cite{zhou2016first}. Accordingly, we have analyzed and compared the electrical transport properties obtained using CRTA with those obtained via both EPI-based RTAs. In this study, the transport properties under the CRTA are estimated using the BoltzTraP code \cite{madsen2006}. To enable a clear comparison between the CRTA and EPI-based methods, the required electron-energy independent lifetime in CRTA is set equal to the value obtained from \eph\ scattering (here, taken from MRTA) at a given T. In this context, the effects of \eph\ scattering, or equivalently carrier lifetimes, are reflected through variations in carrier concentrations in the SERTA or MRTA approaches. All three electrical transport coefficients are calculated using the \textit{G$_0$W$_0$} band gap values (applied via scissor correction), which have been estimated in our previous study \cite{solet2025many} to be approximately 1.5 eV for \lz\ and 1.0 eV for \sa. Given that the \mue\ is approximately 80–100 times higher than the \muh\ (see Fig.~\hyperref[fig:mob]{\ref{fig:mob}}), our investigation focuses only on the n-type case to research their TE properties, where enhanced efficiency is anticipated.

\begin{figure}[ht]
\includegraphics[width=7.5cm, height=9.0cm]{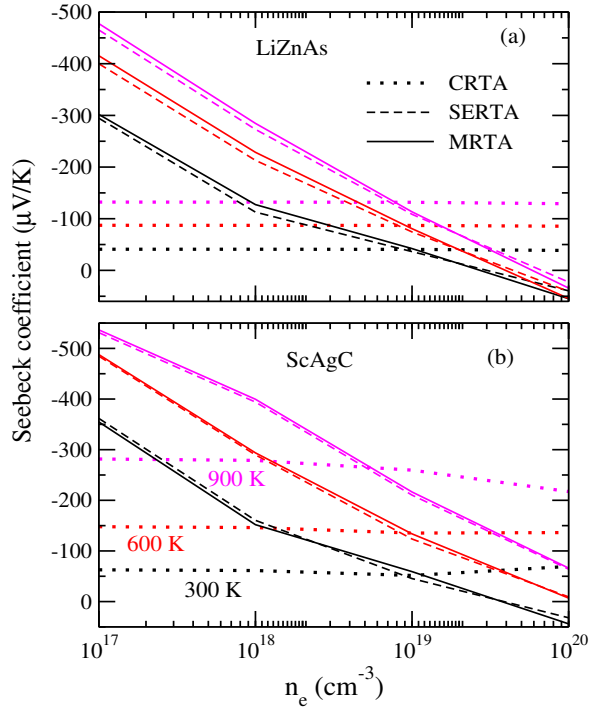} 
\caption{The Seebeck coefficients of n-type of (a) \lz\ and (b) \sa\ hH materials at different temperatures and increasing electron doping concentrations (n$_e$), calculated using the CRTA, SERTA and MRTA approaches.} 
\label{fig:sbk}
\end{figure}

We begin by analyzing the Seebeck coefficients (S) for both n-type hHs calculated using the CRTA, SERTA, and MRTA methods, which are plotted in Fig.~\hyperref[fig:sbk]{\ref{fig:sbk}} as a function of n$_e$ at 300, 600, and 900 K. It is clearly demonstrated for both hHs that the physical behavior of S appears counterintuitive when relying solely on the CRTA framework. However, both SERTA and MRTA provide nearly similar S values, indicating that the average change of electron velocity during scattering processes in Eq.~\hyperref[eq:tau]{\eqref{eq:tau}} is negligibly small. The magnitude of S (|S|) increases with T and decreases with increasing n$_e$ at a given T, analogous to a typical behavior of most semiconducting TEs. Since lifetime is assumed to be constant in CRTA, the value of S is not significantly improved with increasing n$_e$, which can also be due to the large band gap of these hHs. Inclusion of \eph\ scattering leads to a noticeable improvements in S across all T when compared to CRTA estimates, with the most significant enhancements observed at lower n$_e$. From the band structure, intraband scattering is expected to play an important role in producing these higher S values. Specifically, at a n$_e$ of 10$^{17}$ cm$^{-3}$, S values are 2-6 times higher than those predicted by CRTA in both hHs. However, this discrepancy decreases as n$_e$ increases. 

Furthermore, S at the CRTA level is estimated to be always negative for both n-types \lz\ and \sa, in line with conventional expectations. The EPI-based results, which take into account the \eph\ scattering within the SERTA (or MRTA), reveal a sign reversal in S, becoming positive at large enough electron doping levels (10$^{20}$ cm$^{-3}$) for \lz\ at all T, and for \sa\ at 300 K. This type of sign-change behavior at high doping concentrations has also been observed in other hHs, such as in TaFeSb \cite{fedorova2022anomalous}. This unusual trend arises from carrier scattering events between various electronic states, governed by their energy-dependent interactions. This highlights the importance of incorporating advanced scattering mechanisms to accurately capture carrier transport in typical TE materials \cite{fedorova2022anomalous}. The calculated S values are higher in \sa\ than in \lz, which can be attributed to the less dispersive electronic bands in \sa\ compared to those in \lz. For instance, the |S| for \lz\ at n$_e$ $\sim$ 10$^{18}$ cm$^{-3}$ are in the range of 113 to 285 $\mu$VK$^{-1}$ within both EPI-based RTAs, as T increases from 300 to 900 K. Whereas \sa\ exhibits a higher |S| range of 160 to 400 $\mu$VK$^{-1}$ over the same T interval. This high-T value for \sa\ is comparable to that of the highly efficient TaFeSb hH \cite{fedorova2022anomalous}. It has been observed in many n-type hHs that high $zT$ is typically associated with S in the range of -200 to 250 $\mu$VK$^{-1}$ \cite{rogl2023development}. Therefore, both \lz\ and \sa\ demonstrate promising characteristics for mid- and high-T TE applications.

\begin{figure}[ht]
\includegraphics[width=7.0cm, height=9.0cm]{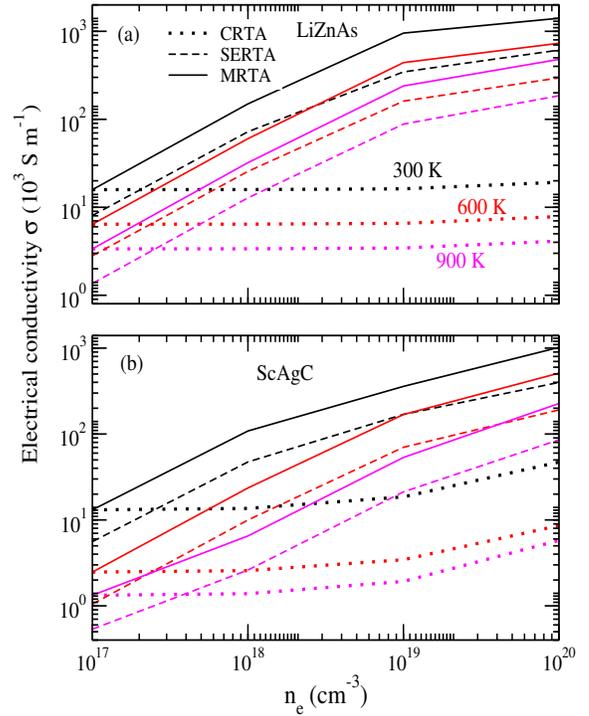} 
\caption{Electrical conductivity of n-type of (a) \lz\ and (b) \sa\ hH materials at different temperatures and increasing electron doping concentrations (n$_e$), calculated using CRTA, SERTA and MRTA.} 
\label{fig:sigma}
\end{figure}

Next calculated $\sigma$ is shown in Fig.~\hyperref[fig:sigma]{\ref{fig:sigma}}. It is observed to show a decreasing trend of $\sigma$ with increasing T at a fixed concentration, consistent with the change of \mue\ with T, which is expected due to enhanced \eph\ scattering at higher T. Due to the electron energy-independent lifetime in CRTA, the $\sigma$ value does not significantly improve with increasing n$_e$, as seen in both hHs. However, after considering the electronic state–dependent lifetime $\tau_{n\mathbf{k}}^{0}$ within the EPI framework, $\sigma$ increases linearly with rising n$_e$ at a fixed T, despite the decreasing \mue\ [see Eq.~\hyperref[eq:serta]{\eqref{eq:serta}}], likely due to the enhanced electronic density of states around the CBM region. This behavior follows the relation of electron relaxation time with T, as well as the general trend $\sigma$ $\propto$ n$_e$. Thus, the strong energy dependence of lifetime reflects to a significant enhancement of $\sigma$ at each T as doping density increases. At fixed T and n$_e$ values, the MRTA-derived $\sigma$ values are higher than those estimated using the SERTA, indicating that the relative change in electron velocity during scattering cannot be neglected in the computation of $\sigma$. Finally, the highest obtained $\sigma$ range of $\sim$32125-479942 Sm$^{-1}$ for \lz\ using MRTA at 900 K across n$_e$ range of 10$^{18}$-10$^{20}$ cm$^{-3}$ is higher than that of \sa\ ($\sim$6500-226500 Sm$^{-1}$), as expected due to the larger \mue\ in \lz. Compared to other potential hHs at high-T \cite{rogl2023development}, these results clearly indicate the potential for achieving high TE performance in both n-type \lz\ and \sa.

\begin{figure}[ht]
\includegraphics[width=7.0cm, height=9.0cm]{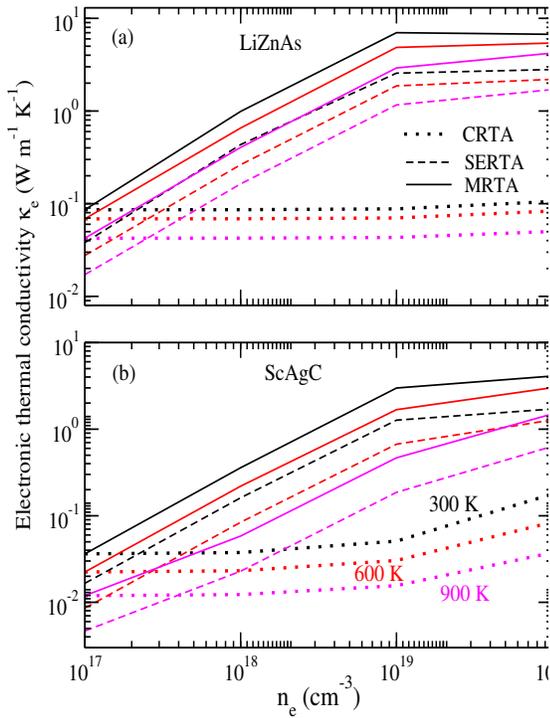} 
\caption{Electronic thermal conductivity of n-type of (a) \lz\ and (b) \sa\ hH materials at different temperatures and increasing electron doping concentrations (n$_e$), calculated using the CRTA, SERTA and MRTA approaches.} 
\label{fig:ke}
\end{figure}

The calculated $\kappa_e$ for both n-type hHs under three RTAs are plotted in Fig.~\hyperref[fig:ke]{\ref{fig:ke}}, in which a similar trend of plots to that observed in $\sigma$ is evident. A slight variation in the CRTA-predicted values with increasing n$_e$ for all T values is seen for both hHs. However, a large enhancement in $\kappa_e$ occur when EPI scattering is included, as clearly observed in $\kappa_e$ plots. The increase in $\kappa_e$ with n$_e$ is a similar observation that has been observed in experimental studies. This marked change underscores the crucial role in between doping concentration and scattering interactions, which are inherently absent in the constant lifetime method. In the case of \lz, SERTA- and MRTA-based all plots exhibit an initially linear increase in $\kappa_e$ with n$_e$ up to 10$^{19}$ cm$^{-3}$, followed by a modest enhancement at high doping levels. However, for \sa, $\kappa_e$ shows a more linear relationship with increasing T across the studied n$_e$ range. It is also observed that the $\kappa_e$ values from MRTA are more than twice as large as those calculated using SERTA. In general, lattice vibrations control most of the heat flow in semiconducting TE materials, therefore it is very crucial to understand the lattice transport for fully design and optimization of TE materials. 

\begin{figure}[ht]
\includegraphics[width=7.0cm, height=5.5cm]{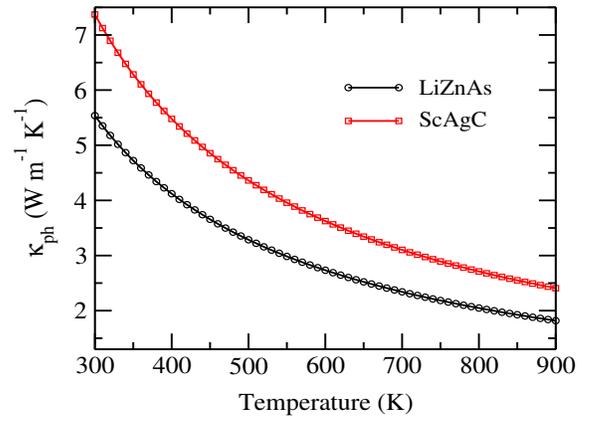} 
\caption{Temperature-dependent lattice thermal conductivity (\kph).} 
\label{fig:kph}
\end{figure}

Therefore, the T-dependent \kph\ calculated from PPI is illustrated in Fig.~\hyperref[fig:kph]{\ref{fig:kph}}. The energy gap between the highest acoustic and the lowest optical branches is smaller in \lz\ than in \sa\ (see the phonon dispersions in Sec.~\ref{sec:3a}), which facilitates stronger phonon-phonon scattering processes, including acoustic–optical interactions within PPI. This enhanced scattering in \lz\ could lead to a lower \kph\ compared to \sa, as also observed in \kph\ plot. The calculated room-temperature values are $\sim$5.5 and 7.4 Wm$^{-1}$ K$^{-1}$ for \lz\ and \sa, respectively. These values are further reduced as T increases due to enhanced phonon-phonon scattering, reaching $\sim$1.8 and 2.4 Wm$^{-1}$ K$^{-1}$, respectively, at 900 K. The calculated values for both hHs at both temperatures are very similar to those of many high-performance hH TEs \cite{rogl2023development}, however they are also lower than those of many other hH materials (room-temperature \kph\ between 10-35 Wm$^{-1}$ K$^{-1}$ \cite{rogl2023development,gandi2017thermoelectric,shastri2020first}).

\begin{figure}[ht]
\includegraphics[width=7.0cm, height=8.5cm]{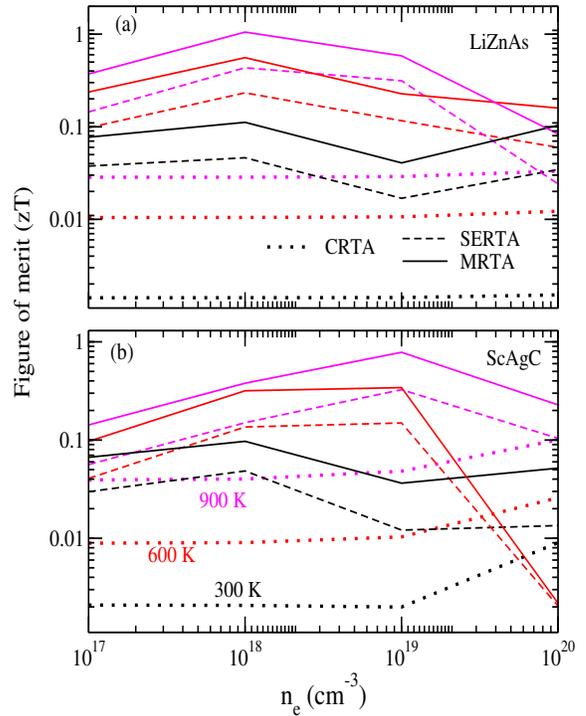} 
\caption{Figure of merit ($zT$) of n-type of (a) \lz\ and (b) \sa\ hH materials at different temperatures and increasing electron doping concentrations (n$_e$), calculated using CRTA, SERTA and MRTA.} 
\label{fig:zt}
\end{figure} 

Now based on the calculated electron transport coefficients and \kph, the TE performance ($zT$) for both hHs has been obtained and is presented in Fig.~\hyperref[fig:zt]{\ref{fig:zt}}. In this figure, one can clearly observe significant differences in the $zT$ values derived from the CRTA and EPI-based approaches. For example, the highest $zT$ for \lz\ is obtained as $\sim$1.05 and 0.44 using MRTA and SERTA, respectively, at a T value of 900 K for a 10$^{18}$ cm$^{-3}$ electron doping. In comparison, the corresponding values for \sa\ are $\sim$0.78 and 0.33 at a similar T but for a 10$^{19}$ cm$^{-3}$ doping concentration. These values are almost 15-35 times larger than the CRTA-based estimations for \lz\ and around 7-16 times higher in \sa. This large discrepancy in the $zT$ results arises mainly from significantly lower S and $\sigma$ values predicted by the CRTA method. Although the $\kappa_e$ is lower within CRTA, this reduction does not sufficiently compensate the former reduction, resulting in overall lower $zT$ values compared to those from EPI-based methods. This comparative analysis clearly highlights the significant impact of methodological choices on accurate $zT$ predictions, underscoring the importance of careful consideration when interpreting and applying these results to the final performance evaluation of hH materials.

As discussed in the introduction section, due to the interrelation between electron transport coefficients, reducing \kph\ is a more efficient way to increase the performance of TE materials. This can be reduced by controlling the lattice transport without largely affecting the electron transport. The grain boundaries are commonly found in experimental samples and thus act as additional sources of scattering centers. Therefore, an effective approach to modulate transport behavior in materials is nanostructuring, which affects transport by causing scattering of carriers at grain boundaries \cite{qiu2015first,minnich2009modeling,dresselhaus2007new,joshi2011enhancement}. The transport of both particles significantly depends on the size of grain or nanostructure. The grain size can be selected by understanding the mean-free-path (MFP) distributions of both types of particles. The electron MFPs are much shorter than phonon MFPs in semiconductors \cite{qiu2015first}, thus nanograins of moderate size can effectively block lattice heat conduction while having minimum impact on electrical transport. This approach has been seen for lowering \kph\ in semiconductors when the size of nanostructures is chosen in the range of 10-100 nanometers (nm) \cite{qiu2015first,chen2004encyclopedia}.

\begin{figure}[ht]
\includegraphics[width=6.5cm, height=8.5cm]{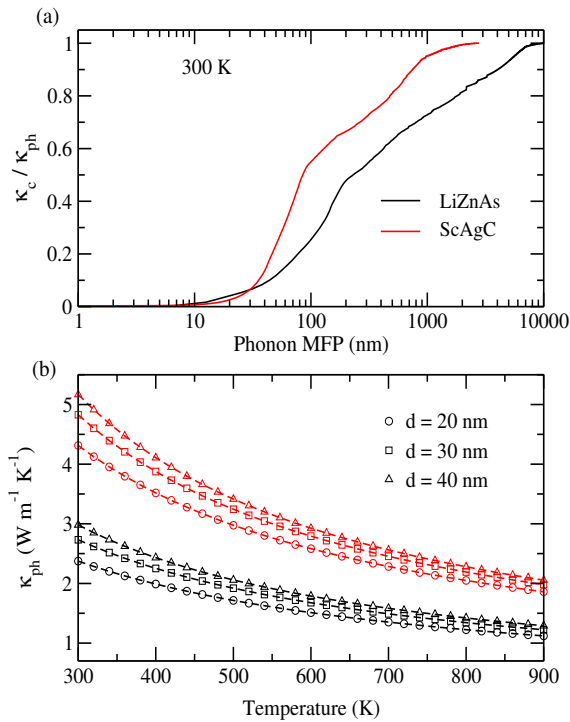} 
\caption{(a) Normalized cumulative lattice thermal conductivity ($\kappa_c$/\kph) at room-temperature as a function of phonon mean-free path (MFP). (b) Temperature dependence of \kph\ by also including phonon boundary scattering using several grain sizes d = 20, 30, and 40 nm.} 
\label{fig:kphb}
\end{figure} 

To support the proper selection of nanograin size, the room-temperature cumulative lattice thermal conductivity is analyzed and presented in Fig.~\hyperref[fig:kphb]{\ref{fig:kphb}(a)}, normalized by the bulk \kph\ value ($\kappa_c$/\kph) for these hH materials. It can be observed for both materials that phonons with MFPs below 10 nm contribute very little to heat conduction at 300 K. However, \kph\ receives significant contributions from phonons with MFPs upto almost 3 $\mu$m (for \sa) and 9 $\mu$m (for \lz). It has been observed in most semiconductors that the large part of electron transport properties such as \mue\ and $\sigma$ \cite{li2015electrical,qiu2015first,solet2025band} are decided by electron with MFPs shorter than 10 nm. This suggests that one can safely choose the grain size above 10 nm in both hH materials, as it would effectively scatter phonons without significantly affecting electron transport.

Therefore, we have selected various grain sizes to study the effect of nanostructuring on the \kph\ value. In this study, the nanograins size d = 10, 20, and 30 nm have been chosen, and the resulting \kph\ plots for both nanograined hH materials are shown in Fig.~\hyperref[fig:kphb]{\ref{fig:kphb}(b)}. This nanoinclusions effect in the calculation is considered in terms of the phonon-boundary scattering, using boundary scattering rates as v$_g$/d, where v$_g$ is the phonon group velocity and d is the boundary MFP ( or grain size). For both materials, the bulk \kph\ value decreases as the d value is reduced from 40 nm to 20 nm, which is expected due to enhanced boundary scattering at smaller grain size. For instance, the room-temperature bulk \kph\ for \lz\ of 5.5 Wm$^{-1}$ K$^{-1}$ reduces to be $\sim$3, 2.7, and 2.4 Wm$^{-1}$ K$^{-1}$ for d values of 40, 30, and 20 nm, respectively. Similarly, the value of 7.4 Wm$^{-1}$ K$^{-1}$ for \sa\ is reduced to be $\sim$5.2, 4.8, and 4.3 Wm$^{-1}$ K$^{-1}$ for the respective d values. However, in both materials, the differences in \kph\ values among these d values become very little at higher T, indicating that phonon-boundary scattering becomes less effective as T increases. 

Using these nanograined sample values, we further calculate the $zT$ values. A strong impact of nanostructuring is observed on these hH materials' performance. For example, the highest $zT$ for \lz\ obtained at 900 K and 10$^{18}$ cm$^{-3}$ concentration, 1.05 using MRTA, largely increases to $\sim$1.53, 1.44, and 1.38 for d values of 20, 30, 40 nm, respectively. At these respective d values, the increase from the value of 0.78 for 10$^{19}$ cm$^{-3}$ concentration in \sa\ is observed to be $\sim$1.0, 0.92, and 0.9. These results indicate that, due to nanostructure technique, the highest $zT$ value at high-T is increased by almost 50\% in \lz\ and 25\% in \sa\ with a 20 nm nanograin size. Such a substantial enhancement shows the major significance of nanostructuring techniques in predicting high-performance TE materials. 

It is very interesting to note that in this work the \kph\ is estimated by considering only three-phonon scattering. However, in recent years, the inclusion of higher-order scattering processes, such as four-phonon scattering~\cite{han2022fourphonon}, has been shown to significantly reduce \kph\ further. For example, this effect on \kph\ in several hH TEs~\cite{yang2024hh130} has been investigated, and reductions of more than 40\% have been reported in some compounds. Also, this scattering mechanism becomes more pronounced at high temperatures. Therefore, one can expect a significant reduction in \kph\ for the hH materials studied here as well, which may lead to very large $zT$ values in nanostructured samples. Testing the effect of four-phonon scattering would thus be a valuable direction for future work.

Finally, in our previous study \cite{solet2025many}, both hH materials have been predicted to be good solar cell materials, exhibiting high efficiencies (greater than 30\%) in single-junction solar cells. Since the \lz\ compound has already been prepared experimentally \cite{solet2025many}, the successful fabrication of \sa\ could also prove beneficial for energy-related applications. Therefore, both hH materials hold promise as multifunctional candidates, capable of efficient performance in both solar cell and TE technologies.

\section{Conclusions} \label{sec:concl}

In summary, we have employed comprehensive \emph{ab initio} MBPT calculations based on DFT and DFPT to investigate the impact of \eph\ coupling on the electronic band structure, electrical carrier mobilities, and other charge transport coefficients of \lz\ and \sa\ hH compounds. All relevant quantities, including electronic and phononic eigenvalues, and \eph\ matrix elements are estimated from first principles. The ZPR corrections of $\sim$15-16 and 36-39 meV to the KS band gap value of $\sim$0.6 and 0.45 eV in \lz\ and \sa, respectively, has been estimated using non-adiabatic AHC calculations. Furthermore, the electron transport properties are computed using BTE with various RTAs, including SERTA and MRTA as well as IBTE under EPI calculations. The \muh\ values are significantly lower than the \mue\ values over the T range of 300-900 K, indicating the significant \eph\ scattering around topmost VB region. All three electrical transport components obtained using CRTA over the n$_e$ range of 10$^{17}$-10$^{20}$ cm$^{-3}$ are significantly influenced by the energy-dependent lifetimes within SERTA and MRTA, highlighting the dominant role of \eph\ coupling in these hHs. Furthermore, the \kph\ calculated using PPI is decreased through nanostructuring techniques, modeled via phonon-boundary scattering with varying grain sizes. Finally, $zT$ is obtained to be maximum with value of $\sim$1.05 (0.78) for 10$^{18}$ (10$^{19}$) cm$^{-3}$ at 900 K within MRTA for \lz\ (\sa), and this value is increased to be $\sim$1.53 (1.0) when a 20 nm nanograin is considered. These results clearly indicate the critical role of intrinsic \eph, phonon-phonon, and phonon-boundary scattering mechanisms for accurately obtaining the highly efficient hH TE materials.     

\vspace{0.05in}
\section*{Acknowledgement}
We acknowledge the computational support provided by the High-Performance Computing (HPC) PARAM Himalaya at the Indian Institute of Technology Mandi.

\vspace{0.05in}
\section*{DATA AVAILABILITY}
The data that support the findings of this article are openly available \cite{vinod2026_18708380}. 
	
	

\bibliography{MS6}

\end{document}